\documentclass[12pt]{article}
\usepackage{amssymb,amsmath,epsfig}
\begin{document}
\parindent=0pt
\parskip=6pt
 \rm

\begin{center}

{\large \bf THE EXISTENCE OF A STABLE NONCOLLINEAR PHASE IN A HEISENBERG MODEL
WITH COMPLEX STRUCTURE}

\vspace{0.5cm}

{\bf Diana V. Shopova}$^{\dag,}$$^{\ast}$, and {\bf Todor L.
Boyadjiev}$^{\ddag}$

$^{\dag}${\em CPCM Laboratory, G. Nadjakov Institute of Solid State Physics,\\
Bulgarian Academy of Sciences, BG-1784 Sofia, Bulgaria.}

$^{\ddag}${\em Faculty of Mathematics and Computer Science, University of
Sofia,\\ BG-1164 Sofia, Bulgaria.}

\end{center}
$^{\ast}$ {\em Corresponding author}: email: sho@issp.bas.bg
\vspace{0.3cm}


{\bf PACS number(s)}:75.10.Hk, 75.30.Kz, 75.40.Cx

\begin{abstract}
 We have analyzed the properties of a noncollinear magnetic
phase obtained in the mean-field analysis of the model of two coupled
Heisenberg subsystems. The domain of its
existence and stability is narrow and depends on the ratio between the averaged
over nearest neighbours microscopic exchange parameters.

 \end{abstract}

The study of Heisenberg magnets with complex structure is important for the
understanding of magnetic properties of real substances~\cite{fh:rmp}. In
this letter we
shall present some results of the mean-field analysis based on the model
of two bilinearly coupled classical Heisenberg
subsystems. The proper identification of the possible magnetic phases, their
domain of existence and stability can be described by this fully isotropic
model for a big class of magnetic materials, in which the different types of
anisotropies are several orders smaller than the exchange interaction. The
importance of purely exchange models for the description of phases and phase
transitions in magnetic substances is well elucidated, for example,
in~\cite{von:mag},~\cite{inp:jmmm}.

Our analysis is done with the help of the following microscopic hamiltonian:
\begin{equation}
H =
- \frac{1}{2}
\sum_{ij}^{2N} \left [ J_{ij}^{(1)} {\bf S}_{i}^{(1)} \cdot {\bf S}_{j}^{(1)} +
J_{ij}^{(2)} {\bf S}_{i}^{(2)} \cdot {\bf S}_{j}^{(2)} +
2 K_{ij} {\bf S}_{i}^{(1)} \cdot {\bf S}_{j}^{(2)}  \right ] .
\end{equation}
The variables ${\bf S}_{i}^{\alpha}$ are $n$-component classical spin vectors
which obey the condition $\left | {\bf S}_{i}^{(\alpha)} \right |  = 1$;
($\alpha = 1, 2$). The dimensionless parametres,
\begin{eqnarray}
J_{ij}^{(\alpha)}& = & \frac{{\cal{J}}_{ij}^{(\alpha)}}{T}\:,
\\ \nonumber
K_{ij}&  = & \frac{{\cal{K}}_{ij}}{T}\:, \nonumber
\end{eqnarray}

where $T$ is the temperature, stand for the in-subsystems  and intersubsystem
exchange interactions,
respectively. In the general case they are elements of $(\mbox{N $\times$
N})$ matrix with $N$ - the number of lattice sites in the subsystems.

The
Ginzburg-Landau functional for the hamiltonian (1) is obtained in our previous
paper
~\cite{dt:2001}
with the help of the Hubbard-Stratonovich transformation
(HST).
Somewhat different way of application of HST to the model (1) can be found in
the book~\cite{ilf:fef}, where a  RG analysis of the same model is done.

In the mean field calculations  we neglect the spatial dependence of the order
parameters
and obtain the Landau free energy density in the
form,
 \begin{eqnarray}
g  \equiv  \frac{G}{NT} &=&
\frac{\tau_1}{2}{\mbox{\boldmath$\varphi$}}^2_1
+\frac{\tau_2}{2}{\mbox{\boldmath $\varphi$}}^2_2
+\frac{g_1}{4} \left( \mbox{\boldmath $ \varphi$}^2_1 \right)^2
+\frac{g_2}{4} \left( \mbox{\boldmath $ \varphi$}^2_2 \right)^2
  +\frac{v}{2}\mbox{\boldmath $ \varphi$}^2_1\mbox{\boldmath $
\varphi $}^2_2   \\ \nonumber
&& +v \left(\mbox{\boldmath $ \varphi$}_1 \cdot \mbox{\boldmath $\varphi$}_2
\right)^2 +w_1 \mbox{\boldmath $ \varphi$}^2_1 \left(\mbox{\boldmath $
\varphi$}_1 \cdot \mbox{\boldmath $ \varphi$}_2
\right) +w_2 \mbox{\boldmath $ \varphi$}^2_2 \left(\mbox{\boldmath $
\varphi$}_1 \cdot \mbox{\boldmath $ \varphi$}_2 \right) \; . \nonumber
\end{eqnarray}
The $n$-component real-space vectors
$\mbox{\boldmath$\varphi$}_i, \;  (i=1,2)$,
play the role of order parameters of the system.
 We shall consider only
a positive in-subsystem exchange and the weak-coupling
limit
which means that the interaction between the subsystems is smaller than the
in-subsystem
interactions.  As far as we do not take into account any magnetic anisotropies,
we shall limit our considerations to magnetic substances with a cubic crystal
structure
as the bcc  lattice with two different magnetic ions per unit cell. Under these
conditions, the model (1) will describe two interpenetrating ferromagnetically
ordered
sublattices which interact either ferromagnetically or antiferromagnetically.

 For a ferromagnetic in-subsystem exchange the coefficients in Eq.~(3) will be
expressed by the quantities
$S_0
(\mbox{\boldmath$k$}),
\;  S_1
(\mbox{\boldmath$k$})$
and $\lambda_{1,2}
(\mbox{\boldmath$k$})
$ for
$\mbox{\boldmath$k$}
= 0$, where
$S_0(\mbox{\boldmath$k$})
\; \mbox{and}\;
S_1(\mbox{\boldmath$k$})
$,
are the elements of the
unitary
matrix, $\hat{S}$, which diagonalizes the Ginzburg-Landau functional in
$\mbox{\boldmath$k$}-$space
and $\lambda_{1,2}(\mbox{\boldmath$k$})
$
are the respective eigenvalues.
The matrix
elements and eigenvalues are functions of the averaged
over the  nearest neighbours microscopic exchange parameters (2)
denoted by $J_1,\; J_2, \; K$, respectively, for details, see~\cite{dt:2001}.
The Landau free energy depends on the
number of order parameter components $n$ through the
parameter $u =1/n^2(n+2)$ which enters linearly in all coefficients in front
of the fourth order terms. The coefficients in front of the second order terms
$\tau_{1,2}$ are also functions of $n$;
$\tau_{1,2}=1-\lambda_{1,2}/n$.

The initial hamiltonian, Eq.~(1), is degenerate with respect to the rotations
of all spins
by one and the same angle and this  degeneracy is preserved in the Landau
free energy density, Eq.~(3). That is why from the mean-field equations:

$$\frac{\partial g}{\partial \varphi_{1i}} = 0\;, \;\;\;\;\;\;\;\;
\frac{\partial g}{\partial \varphi_{2i}} = 0, \;\;\;\;\;\;\;\;
(i=1,\ldots,n)\;,$$

we can find
for $n>1$,
only the magnitudes of the vector order parameters
$\mbox{\boldmath$\varphi$}_i, \; (i=1,2)$
 and their mutual orientation. Our previous
analysis shows that concerning the mutual orientation of vector order
parameters, there are two possibilities, namely:
\begin{eqnarray}
\cos{(\mbox{\boldmath$\varphi_1$},\mbox{\boldmath$\varphi_2$})}&=&\pm 1\; -
\;\mbox{collinear phases}\\\nonumber
\cos{(\mbox{\boldmath$\varphi_1$},\mbox{\boldmath$\varphi_2$})}&\neq& \pm 1\; -
\;\mbox{noncollinear phases}.
\end{eqnarray}

In the collinear case the system of equations for the magnitudes of vector
order parameters
$\mbox{\boldmath$\varphi$}_i,  \; (i=1,2)$
can be solved only numerically and the results are described in details
in~\cite{dt:2001}.

The respective equations for the noncollinear phase can
be treated analytically. Because of the degeneracy of Landau free
energy density in a purely exchange approximation for $n>1$, it is more
appropriate to present $\mbox{\boldmath$\varphi$}_1$  and
$\mbox{\boldmath$\varphi$}_2$ by  the direction cosines
$\gamma_i$
 and
$\alpha_i$
and their magnitudes
$\varphi_1  =  \left( \sum_{i=1}^{n}\varphi^2_{1i} \right)^{1/2},\;
\varphi_2 = \left( \sum_{i=1}^{n}\varphi^2_{2i} \right)^{1/2}$, in the
following way,
 \begin{eqnarray}
  \varphi_{1i} & = & \varphi_1\gamma_i,\qquad\qquad\\\nonumber
  \varphi_{2i}& = &\varphi_2\alpha_i \:.
  \end{eqnarray}

Then the Landau free energy density, Eq.~(3) is rewritten with the help of
Eq.~(5) and afterwards minimized with respect
to  $ \alpha_{i} \;   \gamma_{i}, \; \varphi_1 \; \mbox{and} \; \varphi_2$
under the conditions

$$\sum^{n}_{i=1} \alpha_i^2 = 1, \qquad
 \sum^{n}_{i=1} \gamma_i^2 = 1 \; .$$

The resulting mean field equation
for $\cos{(\mbox{\boldmath$\varphi$}_1, \mbox{\boldmath
$\varphi$}_2)}=\sum_{i=1}^{n}\alpha_i\gamma_i$ will be,
\begin{equation}
2v\varphi_1\varphi_2\sum_{i=1}^{n}\alpha_i\gamma_i + w_1\varphi^2_1 +
w_2\varphi^2_2 = 0\;.
\end{equation}
It is obvious from the above expression that the cosine between
the vector order parameters is well defined for $\varphi_1\ne0$
and $\varphi_2\ne0$. The vector order
parameters will be mutually
perpendicular
when $w_1=0,\;w_2=0$. This is possible for equivalent sublattices because
$w_{1,2}\sim(J_1-J_2)$, but in this case the system of equations for $\varphi_1
$ ans $\varphi_2$ has a solution only when $K \equiv 0$, which means fully
decoupled sublattices.

When $J_1 \ne J_2$ the system of equations for $\varphi_1, \; \varphi_2$ is
simple and can be solved directly,
\begin{eqnarray}
\tau_1 +
(g_1-\frac{w_1^2}{v})\varphi_1^2+(v-\frac{w_1w_2}{v})\varphi_2^2&=&0\\\nonumber
\tau_2 + (g_2-\frac{w_2^2}{v})\varphi_2^2+(v-\frac{w_1w_2}{v})\varphi_1^2&=&0
\;.
\end{eqnarray}

To understand the properties of the obtained magnetic phase we need the
connection between the order
parameter vectors $\mbox{\boldmath$\varphi$}_i, \;  (i=1,2)$, and
the sublattice magnetizations,
$\mbox{\boldmath$m$}_i, \;  (i=1,2)$
\begin{eqnarray}
\mbox{\boldmath$m$}_1&=&
\frac{S_0}{\lambda_1} \mbox{\boldmath$\varphi$}_1 -
\frac{S_1}{\lambda_2} \mbox{\boldmath$\varphi$}_2\:,\\\nonumber
\mbox{\boldmath$m$}_2&=&
\frac{S_1}{\lambda_1} \mbox{\boldmath$\varphi$}_1 +
\frac{S_0}{\lambda_2} \mbox{\boldmath$\varphi$}_2 \:.
\end{eqnarray}
 From the above expression we can find
the magnitudes of the sublattice
magnetizations and
the angle $\gamma$ between them.
The requirements
$|\cos{(\gamma)}| \le 1 \; \mbox{and}\; |\sum_{i=1}^n\alpha_i\gamma_i| \le 1 $
restrict the values of parameters $J_1, J_2\; \mbox{and}\;
K$, for which the noncolliner phase can exist.

The analytical expressions
for magnitudes of the sublattice magnetizations and $\cos{(\gamma)}$ can be
obtained straightforwardly  after some algebra but they are very cumbersome,
moreover, the temperature dependence of $m_1, \; m_2$ and $\cos{(\gamma)}$
cannot
be comprehended directly from formulae. For this reason we prefer to give a
graphical presentation  of our results. In order to compare the properties of
the noncollinear and collinear phases, obtained previously,
we shall rewrite the solutions for the noncollinear phase with
the
help of the dimensionless parameters, introduced in~\cite{dt:2001}:
$$x=\frac{T}{{\cal{J}}_1+{\cal{J}}_2}
\;.$$
$$\alpha=\frac{{\cal{J}}_1-{\cal{J}}_2}{{\cal{J}}_1+{\cal{J}}_2}, \qquad
 \beta=\frac{2{\cal{K}}}{{\cal{J}}_1+{\cal{J}}_2} \;.$$
The quantity $x$ is called the reduced temperature; $\alpha$ can be considered
as a measure of the difference between the two
sublattices.
We assume that, ${\cal{J}}_1>{\cal{J}}_2$, because the
sublattices
are symmetric with respect to the interchange $1\leftrightarrow 2$; then,
$0<\alpha<1$.
 The dimensionless intersublattice interaction, $|\beta|<1$, which is
a requirement of the weak-coupling limit ${\cal{J}}_1{\cal{J}}_2-{\cal{K}}^2 >
0$.

In Fig.~1 the variation
of $ \cos{(\gamma)}$ with the reduced
temperature $x$ is depicted for a fixed
value of $\alpha$ and three different values of $\beta$.
 The behaviour
of $\cos{(\gamma)}$  is essentially different  for different values of the
relation
$\alpha / \beta$. When $\alpha\ge \beta$, the
noncollinear phase is metastable in the whole domain of its existence and
$cos{(\gamma)}<1$, including the phase transition point. In the opposite case,
i.e,
$\alpha<\beta$, the noncolliner phase becomes the most stable phase in the
temperature interval $x_n<x\le x^{\ast}, \; \mbox{where} \;x^{\ast}$
is the transition temperature of the noncollinear phase and $x_n$ is the
temperature, below which the noncollinear phase does not exist.
For $\beta=0.003$, the angle
between the sublattice magnetizations vary from zero at $x^{\ast} $ to $\sim
68^{\circ} \; \mbox{at} \; x_n$. When $\beta$ grows (see the curve for
$\beta=0.005$) the domain of the existence for the noncollinear ferrimagnetic
phase becomes more narrow and for $\beta \sim 0.0056$ it does not exist.
\begin{figure}[tbp]
\begin{center}
\epsfig{file=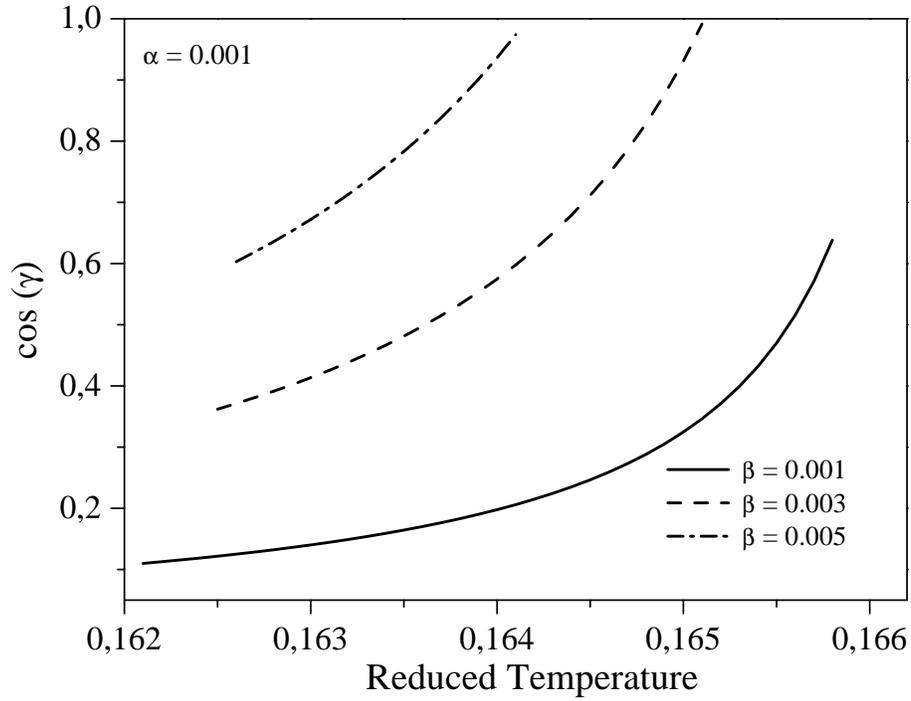, width=12cm}\\
\end{center}
\caption{The temperature dependence of $\cos{(\gamma)}$ for $n =3$,
$\alpha = 0.001$ and different values of the parameter $\beta$.}
\label{F1.fig}
\end{figure}

The calculations show that in the region of stability of the noncollinear phase
the magnetization
$\mbox{\boldmath$M$} = \mbox{\boldmath$m_1$}+\mbox{\boldmath$m_2$}$ is
perpendicular to the staggered magnetization $\mbox{\boldmath$L$} =
\mbox{\boldmath$m_1$}-\mbox{\boldmath$m_2$}$. This is the condition for the
existence and stability of a noncollinear ferrimagnetic phase, derived on the
basis of a symmetry group analysis by Andreev and Marchenko
~\cite{am:1980}.

In Figs.~2 and~3 we compare the Landau free energy density
as a function of the reduced
temperature $x$ for all magnetic phases we have found (collinear and
noncollinear).

\begin{figure}[tbp]
\begin{center}
\epsfig{file=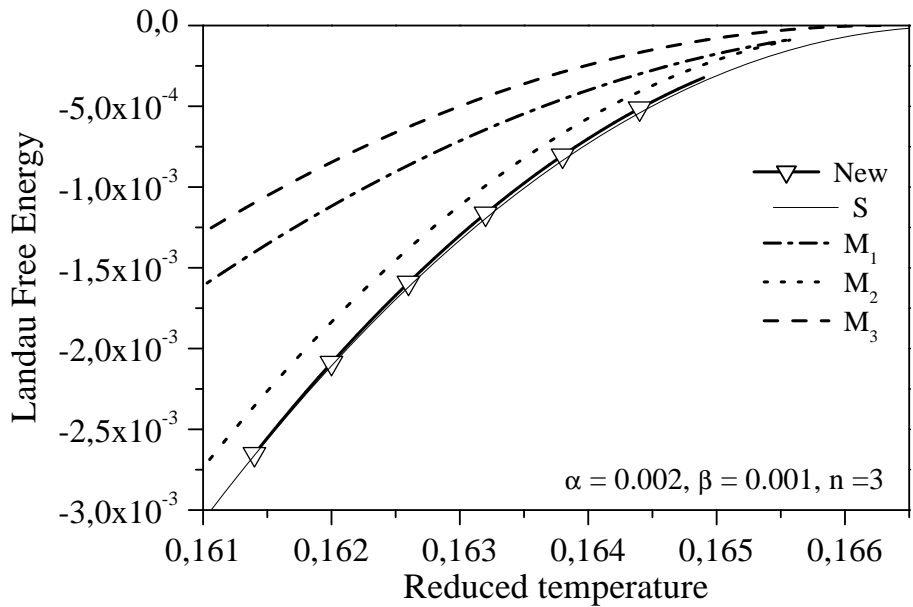, width=12cm}\\
\end{center}
\caption{The temperature dependence of Landau free energy density for
$\alpha>\beta$; the legend is explained in the text.} \label{F2.fig}
\end{figure}

 The  phase with the highest transition temperature, denoted by
$S$ in Figs.~2,~3, has a collinear ferrimagnetic  structure with a compensation
point at lower temperatures. It is stable and occurs through a second order
phase transition. The symbols for the high- and
low-temperature metastable collinear phases are
$M_i,\; i=1,2$, respectively. $M_3$ is a collinear metastable phase, which
occurs by a second order phase transition, not given in~\cite{dt:2001}.
The Landau energy density of the
noncollinear phase is denoted by $New$.

\begin{figure}[tbp]
\begin{center}
\epsfig{file=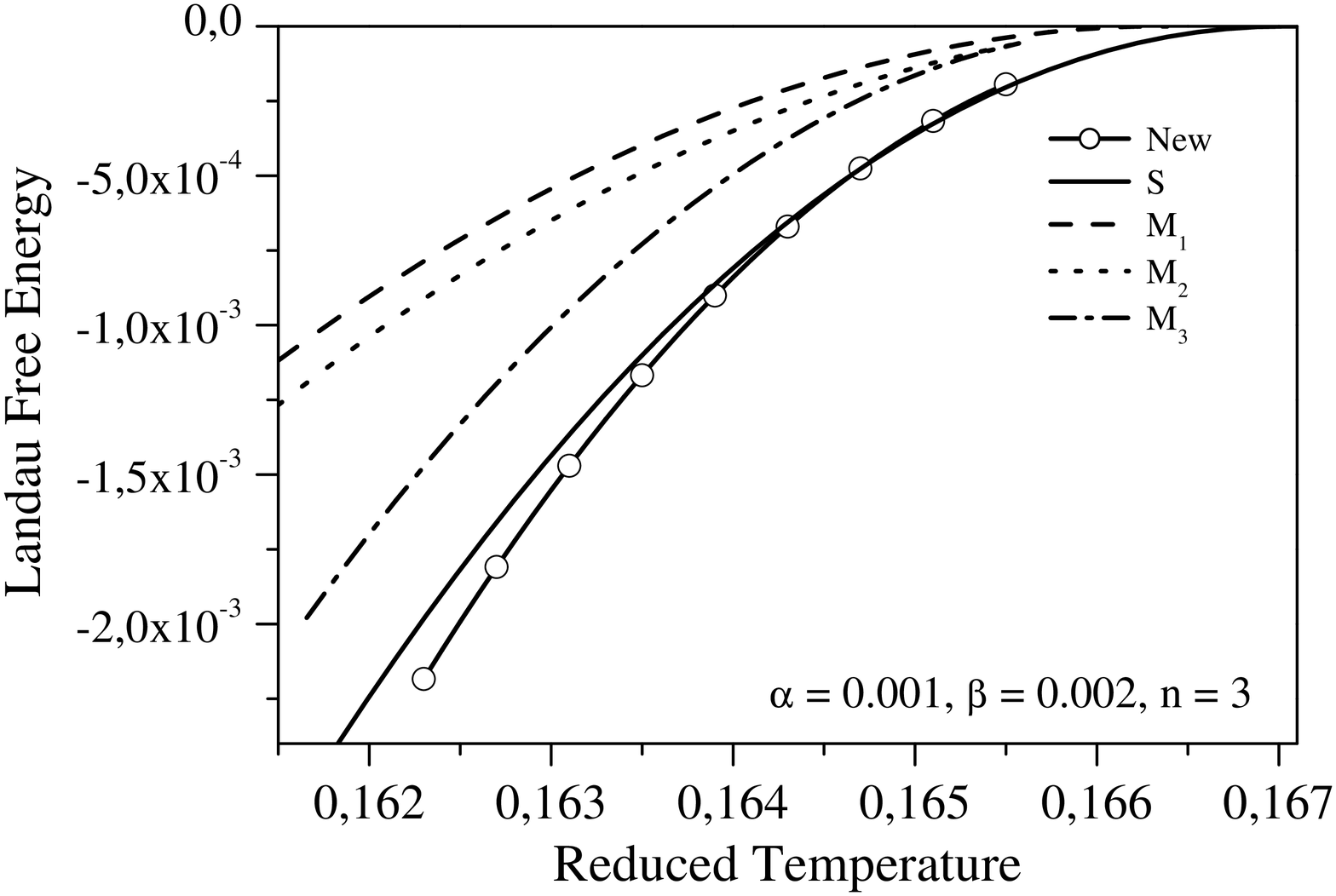, width=12cm}\\
\end{center}
\caption{The temperature dependence of Landau free energy density for
$\alpha<\beta$; the legend is explained in the text.}
 \label{f3.fig}
\end{figure}
For
$\alpha>\beta$ the noncollinear phase has an energy slightly higher than the
stable collinear phase - Fig.~2. When the difference between the interaction
in the sublattices
becomes smaller than the intersublattice interaction, the noncollinear phase
has
lower energy than the $S$-phase in the whole domain of its existence - Fig.~3.

We can argue that with the decrease of the temperature the system orders
by a second order phase transition in the stable collinear ferrimagnetic
phase and
with the further lowering of temperature  a second phase transition occurs and
a new ordered noncollinear phase appeares. At lower temperatures the
noncollinear phase ceases to exist.
It is obvious from the above figures that the domain of existence and stability
of the noncollinear phase is very narrow and it can occur only for
magnetic substances, in which the
following relation between the averaged
microscopic exchange parameters holds true:
$({\cal{J}}_1-{\cal{J}}_2)<{\cal{K}}<\sqrt{{\cal{J}}_1{\cal{J}}_2}$.

Up to now we have considered only the ferromagnetic exchange between the
sublattices ($\beta>0$).
As the calculations show the respective noncollinear phase
for $\beta<0$, i.e., for an antiferromagnetic coupling, exhibits a similar
temperature behaviour as the ferrimagnetic noncollinear phase described above.

We did not study  the noncollinear phase in the general case
of $n$-component order parameters. The above calculations are made for $n=3$
which is the most interesting case up to our opinion. The analysis shows
that for $n=2$ the domain of existence of the noncollinear phase is very
tiny and we did not find values of the parameters $\alpha \; \mbox{and} \;
\beta$, for which it can be stable.

The difference between the energies of the stable collinear ferrimagnetic and
the noncollinear ferrimagnetic phases is very small. In the real magnetic
substances even a small anisotropic interaction may change the above described
picture, but this problem needs a
separate investigation.

{\bf Acknowledgements}

The authors thank Dimo Uzunov for the valuable discussions.

\end{document}